\documentclass[aps,pra,twocolumn,showpacs,superscriptaddress,nofootinbib]{revtex4-2}
\usepackage{amsmath,amssymb,amsfonts}
\allowdisplaybreaks[4]
\usepackage{graphicx}
\usepackage{dcolumn}
\usepackage{epstopdf}
\usepackage[T1]{fontenc}
\usepackage{newtxtext}
\usepackage{subfigure}
\usepackage[percent]{overpic}
\usepackage{float}
\usepackage{bm}
\usepackage{algorithm}
\usepackage{algpseudocode}
\usepackage{booktabs}
\usepackage{float}
\usepackage{booktabs}

\allowdisplaybreaks

\usepackage[unicode]{hyperref}
\hypersetup{
	unicode=true,          
	a4paper=true,
	plainpages=false,
	pdftitle={Title of PDF},    
	pdfauthor={Author of PDF},     
	pdfsubject={Subject of PDF},   
	colorlinks=true,       
	linkcolor=blue,          
	citecolor=blue,        
	filecolor=blue,      
	urlcolor=blue           
}
\newcommand{\refeq}[1]{Eq.~(\ref{#1})}

\newcommand{\reffig}[1]{Fig.~\ref{#1}}
\newcommand{\reffigure}[1]{Figure~\ref{#1}}
\newcommand{\reftable}[1]{Table~\ref{#1}}

\DeclareMathAlphabet{\mathcal}{OMS}{cmsy}{m,b}{n,it}

\begin{document}

\title{Confinement-controlled pattern selection in a finite population-imbalanced dipolar Bose-Einstein condensate}

\author{Zhenhao Wang}
\author{Weijing Bao}
\author{Jia-Rui Luo}
\affiliation{College of Physics, Nanjing University of Aeronautics and Astronautics, Nanjing 211106, China}
\affiliation{Key Laboratory of Aerospace Information Materials and Physics (NUAA), MIIT, Nanjing 211106, China}

\author{Gentaro Watanabe}
\affiliation{School of Physics and Zhejiang Institute of Modern Physics, Zhejiang University, Hangzhou, Zhejiang 310027, China}
\affiliation{Zhejiang Province Key Laboratory of Quantum Technology and Device, Zhejiang University, Hangzhou, Zhejiang 310027, China}

\author{Kui-Tian Xi}
\email[Corresponding author: ]{xiphys@nuaa.edu.cn}
\affiliation{College of Physics, Nanjing University of Aeronautics and Astronautics, Nanjing 211106, China}
\affiliation{Key Laboratory of Aerospace Information Materials and Physics (NUAA), MIIT, Nanjing 211106, China}

\date{\today}

\begin{abstract}
	We study the ground-state density patterns of a population-imbalanced two-component dipolar Bose-Einstein condensate confined in a circular quasi-two-dimensional box. Using a mean-field model, we map out phase diagrams as functions of the axial confinement, interaction imbalance, and population ratio. The system supports a rich sequence of stationary morphologies, including a nearly uniform pancake state, pancake-droplet and ring-droplet coexistence states, droplet arrays, and concentric rings. These patterns show a close structural correspondence to microphase-separated morphologies in diblock-copolymer systems, with the population imbalance acting as an effective volume fraction that selects the pattern topology. Analysis of the density profiles and structure factors reveals that the modulated states possess an intrinsic nonzero characteristic wave vector, which remains essentially unchanged when the box size is varied. We also find that the characteristic pattern spacing scales linearly with the axial confinement length, indicating that the transverse thickness of the condensate controls the effective in-plane length scale. In a finite circular box, this smooth scaling is interrupted by discrete steps, reflecting geometric frustration and the integer locking of the number of rings or droplets. Our results show that box-trapped dipolar mixtures provide a controllable platform for studying finite-size pattern selection and nonlocal microphase formation in quantum fluids.
\end{abstract}

\maketitle

\section{Introduction}

Spontaneous pattern formation driven by the competition between short-range attraction and long-range repulsion is a ubiquitous phenomenon in nature, emerging in systems ranging from magnetic films and ferrofluids to nuclear pasta and soft matter. A paradigmatic example of such self-organization is found in diblock copolymers, where the chemical incompatibility between segments drives macroscopic phase separation, while the covalent bonds impose a long-range constraint. This frustration leads to microphase separation characterized by periodic nanostructures, including lamellae \cite{Lamellar1980Macromolecules}, cylinders \cite{spheres1978Macromolecules}, spheres \cite{cylindrical1980Macromolecules}, and complex networks \cite{Schick1994PRL,Milner1994PRL,Mortensen1994PRL,Hajduk1994Macromolecules}. To describe this universal behavior, T. Ohta and K. Kawasaki established a celebrated free-energy functional \cite{OhtaKawasaki1986,OhtaKawasaki1988,Kawasaki1990,OhtaKawasakifunctional}, which predicts the topology of the ground state based on a single control parameter: the volume fraction of the components.

In the realm of quantum physics, ultracold dipolar gases offer a pristine and highly tunable platform to explore this universal physics. The realization of the first dipolar Bose–Einstein condensate (BEC) with chromium \cite{Pfau2005PRLCr,Pfau2006PRL} introduced long-range, anisotropic dipole-dipole interactions (DDI) that compete with the isotropic short-range contact interaction (CI). Dipolar gases therefore provide a quantum platform for studying length-scale selection in systems with competing local and nonlocal interactions. The extension to binary dipolar condensates \cite{Zhang2015SciRep,Cornell1998PRL} has further enriched this landscape, providing intercomponent coupling and population imbalance as new degrees of freedom to tailor the effective interaction potentials \cite{Lahaye2009RepProgPhys,Baranov2008PhysRep,Chomaz2023RepProgPhys}.

A key advantage of magnetic atoms lies in their extraordinary controllability. Contact interactions can be precisely tuned via Feshbach resonances \cite{Pfau2005PRLFeshbach,Ferlaino2020PRA}, while the DDI can be engineered by rotating magnetic fields \cite{Pfau2002PRL,Benjamin2018PRL}. In the presence of external potentials \cite{Lewenstein2002PRL,Lewenstein2008PRA}, the interplay between trap geometry and interaction anisotropy determines the stationary states, leading to miscible-immiscible transitions \cite{Wieman2008PRL,Malomed2010PRA,Adhikari2012PRA,Bisset2021PRA}, phase separation \cite{Xi2011PRA,Kumar2019PRA}, and various density-modulated phases \cite{Zillich2012PRL,Blakie2024PRA,zhang2025polarization}.

In the regime of strongly magnetic lanthanides such as erbium \cite{Ferlaino2012PRL} and dysprosium \cite{Benjamin2011PRL,Ferlaino2018PRL}, quantum fluctuations encoded in the Lee-Huang-Yang (LHY) corrections \cite{LeeYang1957PR,LeeHuangYang1957PR,Lima2012PRA} play a decisive role, providing the repulsive stabilization necessary for the formation of self-bound quantum droplets \cite{Blakie2021PRL,Santos2021PRL,Saito2022PRA,Zhang2024PRA} and supersolid states \cite{Bisset2022PRL,Santos2023PRA,Majumder2023PRA,Santos2023PRR,Zhang2024PRR}. While these beyond-mean-field effects are essential in free space, the mean-field landscape of confined binary mixtures offers an equally rich, yet distinct, avenue for pattern formation. Here, the interplay of nonlocal interactions and external geometry drives symmetry breaking even at the mean-field level, providing a robust platform for engineering crystalline orders without relying on quantum fluctuation terms. Beyond ground states, this dynamical phase space hosts a plethora of topological textures, including solitons \cite{Santos2009PRL,Malomed2012PRA}, vortices \cite{YiSu2006PRA,Tomio2024PRA,Tomio2025PRA,Simon2005PRL,Kumar2017PRA}, and skyrmions \cite{Zhang2017PRA,Saito2025PRA}, while modulational instabilities \cite{Marklund2010PRA,Saito2012PRA,Xi2018PRA,Blakie2022PRA} offer dynamical pathways to access these emergent structures.

Recent advances in trapping techniques have enabled the confinement of BECs in quasi-uniform optical box potentials \cite{Raizen2005PRA,Hadzibabic2013PRL}. By removing the masking effects of harmonic inhomogeneity, box traps expose the intrinsic competition between the nonlocal DDI and the system boundaries. In this context, the sharp confinement acts not merely as a container but as a source of geometric and commensurability constraints and frustration. While recent studies have probed boundary-induced supersolidity \cite{Recati2022PRR} or vortex turbulence \cite{Tsubota2019PRA,Yakimenko2023PRR,Tsubota2025Turbulence} in these geometries, a systematic framework connecting the \emph{pattern selection} in binary dipolar gases to universal classes of microphase separation remains conspicuously absent.

In this work, we study microphase formation in a population-imbalanced binary dipolar BEC confined in a circular box. By comparing the stationary density patterns with the well-known morphologies of diblock-copolymer systems, we identify a close structural correspondence between the population ratio and the effective volume fraction that selects the pattern topology in the Ohta-Kawasaki picture. Within a mean-field framework, we construct phase diagrams in two parameter planes: axial confinement versus population imbalance, and contact-interaction imbalance versus population imbalance. These diagrams reveal a robust sequence of states, ranging from concentric-ring configurations to droplet arrays, with intermediate coexistence regimes between them. We find that the phase boundaries are organized mainly by the atom-number ratio, while the axial confinement and interaction asymmetry primarily shift the thresholds for density modulation. A central result is the emergence of a characteristic length scale in the modulated states. From the density profiles and their structure factors, we show that the pattern spacing $d$ scales linearly with the axial confinement length, 
\[
d \propto l_z \sim \omega_z^{-1/2},
\] 
indicating that the transverse thickness of the condensate controls the effective in-plane length scale. In the finite circular box, this smooth scaling is interrupted by discrete steps, which reflect geometric frustration and the integer locking of the number of rings or droplets. These results highlight box-trapped dipolar mixtures as a useful platform for studying finite-size pattern selection and nonlocal microphase formation in quantum fluids.

The paper is organized as follows. In Sec.~\ref{SEC:formalism}, we introduce the mean-field model and the numerical method. In Sec.~\ref{SEC:Phase_Diagram}, we present the ground-state phase diagrams and discuss their structural correspondence to microphase-separated states in the Ohta-Kawasaki framework. In Sec.~\ref{SEC:structure factor and scaling analysis}, we analyze the characteristic wave-vector selection and the scaling of the pattern spacing. In Sec.~\ref{SEC:the modulated-unmodulated transition}, we characterize the modulation amplitude across the phase diagram. Finally, Sec.~\ref{SEC:Conclusion} summarizes our results.

\section{Formalism}
\label{SEC:formalism}

We consider a binary dipolar Bose-Einstein condensate (BEC) of $^{52}$Cr atoms at zero temperature. Within the mean-field framework, the system dynamics is governed by the coupled Gross-Pitaevskii equations (GPEs):
\begin{align}\label{eq:GPEs}
	i \hbar \frac{\partial}{\partial t} \Psi_{i} \left(\bm{r}\right)
	= \Bigg[ & -\frac{\hbar^2}{2m} \nabla^2 + V \left( \bm{r} \right) + \sum_{j = 1}^{2} g_{ij} \big| \Psi_{j}\left(\bm{r}\right) \big|^2 \nonumber\\
	& + \sum_{j = 1}^{2} \int U_{ij} \left( \bm{r} - \bm{r'} \right) \big| \Psi_{j} \left( \bm{r'} \right) \big|^2 d \bm{r'} \Bigg] \Psi_{i}\left(\bm{r}\right),
\end{align}
where the indices $i,j=1,2$ denote the two components. The wave function $\Psi_{i}(\bm{r},t)$ is normalized to the respective atom number $N_i = \int d\mathbf{r} \, |\Psi_{i}(\mathbf{r},t)|^{2}$. The short-range contact interactions are characterized by the coupling coefficients $ g_{ij} = 4\pi \hbar^{2} a_{ij} / m $, determined by the intra- and inter-component $s$-wave scattering lengths $a_{ij}$ and the atomic mass $m$.
The long-range, anisotropic dipole-dipole interaction (DDI) is given by 
\begin{equation}
	U_{ij}(\bm{r}) = \gamma \frac{\mu_0 \mu_i \mu_j}{4\pi}
	\frac{1 - 3( \hat{\bm{d}} \cdot \hat{\bm{r}} )^2}{|\bm{r}|^3},
	\label{eq:ddi}
\end{equation}
where $\mu_0$ is the vacuum permeability, $\mu_i$ is the magnetic dipole moment, and the dipoles are polarized along a common axis $\hat{\bm{d}}$ (taken as the $z$-axis), with $\hat{\bm{r}} = \bm{r}/|\bm{r}|$ denoting the relative position unit vector. In the experiment, by rapidly rotating the magnetic field \cite{Pfau2002PRL,Benjamin2018PRL}, the coefficient $\gamma$ can be tuned to directly regulate the strength of DDI.

The ground state wave functions $\Psi_{1}(\bm{r})$ and $\Psi_{2}(\bm{r})$ correspond to the stationary solutions that minimize the total energy functional:
\begin{equation}
	E[\Psi_1, \Psi_2] = \sum_{i=1}^2 \left[ E_{i}^{\text{k}} + E_{i}^{\text{tr}} +\sum_{j=1}^2 \left(E_{ij}^{\text{ci}} + E_{ij}^{\text{ddi}}\right) \right],
	\label{eq:Energy}
\end{equation}
where the individual energy contributions are defined as 
\begin{align}
	E_{i}^{\text{k}} &= \int d\bm{r} \frac{\hbar^2}{2m} |\nabla \Psi_i(\bm{r})|^2, \\
	E_{i}^{\text{tr}} &= \int d\bm{r} V(\bm{r}) |\Psi_i(\bm{r})|^2, \\
	E_{ij}^{\text{ci}} &= \frac{1}{2} \int d\bm{r} g_{ij} |\Psi_i(\bm{r})|^2 |\Psi_j(\bm{r})|^2, \\
	E_{ij}^{\text{ddi}} &= \frac{1}{2} \int d\bm{r} |\Psi_i(\bm{r})|^2 \int d\bm{r}' U_{ij} (\bm{r}-\bm{r}') |\Psi_j(\bm{r}')|^2.
\end{align}
These terms represent the kinetic energy, trap potential energy, contact interaction energy, and dipolar interaction energy, respectively.

We consider a geometry that is quasi-two-dimensional (quasi-2D) in the $x$–$y$ plane and tightly confined along the polarization direction $z$. The external potential is modeled as:
\begin{equation}
	V(\bm{r}) = V^{\text{2D}}(\bm{\rho}) + \frac{1}{2} m \omega_z^2 z^2,
	\label{eq:trap}
\end{equation}
where $\bm{\rho} = (x,y)$ represents the in-plane coordinates. The radial confinement is provided by a circular box potential $V^{\text{2D}}(\bm{\rho})$ of radius $R_0$:
\begin{equation}
	V^{\mathrm{2D}}(\bm{\rho}) =
	\begin{cases}
		V_0, & |\bm{\rho}| \ge R_0, \\
		0,   & |\bm{\rho}| < R_0,
	\end{cases}
\end{equation}
with a barrier height $V_0$ sufficiently large to enforce hard-wall-like boundary conditions.
In the regime where the axial excitation energy $\hbar\omega_z$ far exceeds the chemical potential and thermal energy, the motion in the $z$-direction is frozen into the ground state of the harmonic oscillator, $\phi_0(z) = (\pi l_z^2)^{-1/4} e^{-z^2/2l_z^2}$, where $l_z = \sqrt{\hbar/m\omega_z}$ is the characteristic harmonic length.
Under the ansatz $\Psi_i(\bm{r}) = \psi_i(\bm{\rho})\phi_0(z)$, the system reduces to an effective 2D model where the axial confinement renormalizes the interaction strengths. Specifically, the contact coupling scales as $g_{\text{2D}} = g_{ij}/\sqrt{2\pi}l_z$, and the effective dipolar interaction acquires a momentum-dependent kernel governed by $l_z$. Consequently, $\omega_z$ serves as a critical control parameter: it not only determines the condensate thickness but also tunes the competition between short-range and nonlocal interactions, driving the dimensional crossover and pattern formation discussed in the following sections.

To obtain the ground state solutions of \refeq{eq:GPEs}, we employ a preconditioned conjugate-gradient (PCG) method \cite{SHU2024JCP,Bao2025CG}. This algorithm minimizes the energy functional \refeq{eq:Energy} by directly targeting the energy minimum, offering superior convergence speed and robustness for nonlocal interactions compared to standard imaginary-time propagation. In the static calculations, the convolution integrals associated with the dipolar potential are efficiently evaluated via fast Fourier transforms (FFT) on a discrete grid. This framework allows us to systematically map the phase diagram across the parameter space of atom-number ratio, axial confinement, and interaction imbalance, identifying stationary patterns---ranging from miscible profiles to core-rim structures and droplet arrays.

\section{Phase diagram}
\label{SEC:Phase_Diagram}

In this work, we investigate an antiparallel binary dipolar condensate consisting of two internal states of $^{52}$Cr. Both components share the same atomic mass but possess magnetic moments of opposite sign. Specifically, we assign $\mu_{1} = + 6 \mu_{\text{B}}$ and $ \mu_{2} = - 6 \mu_{\text{B}} $, corresponding to atoms in the $^7S_{3}$ manifold prepared in the Zeeman sublevels $m_J = -3$ and $m_J = +3$, respectively, consistent with pioneering chromium dipolar-BEC experiments \cite{Pfau2005PRLCr,Pfau2006PRL}. Here $\mu_{\text{B}}$ denotes the Bohr magneton. The dipoles of both components are polarized along the $z$ axis, $\hat{\bm{d}} = \hat{\bm{z}}$, resulting in dipole-dipole interactions (DDI) of equal magnitude but opposite sign. This antiparallel configuration effectively isolates the competition between intercomponent dipolar forces and population imbalance within the finite geometry.

The condensate is confined in a circular, quasi-two-dimensional box potential with a fixed radius $R_0 = 20$ $\mu$m and a barrier height $V_0 = 100 \hbar \omega_0$, where $\omega_0 = 2\pi \times 100$ Hz defines the characteristic frequency scale. This setup mimics experimentally realized optical-box traps \cite{Recati2022PRR}. The box potential imposes a sharp boundary in the $(x,y)$ plane, rendering the effects of rotational-symmetry breaking and boundary-induced pattern selection particularly transparent.

To facilitate comparisons across different parameter regimes, spatial coordinates are expressed in units of the harmonic-oscillator length, $ \ell = \sqrt{\hbar / (m \omega_0) } $, rendering the simulation domain and box radius dimensionless. Since column densities are the primary observable in experiments, we visualize the in-plane structures via the two-dimensional column density of each component,
\begin{equation}
	n_i(\bm{\rho}) = \int dz | \Psi_i(\bm{\rho},z) |^2,
	\label{eq:2D_density}
\end{equation}
with $\bm{\rho} = (x,y)$. All density profiles presented herein correspond to $n_i(\bm{\rho})$ within the spatial window $|\bm{\rho}| \le R_0$.

We distinguish thermodynamic phases based on the real-space topology of $n_i(\bm{\rho})$. States preserving continuous rotational symmetry with weak radial variation are classified as miscible or nearly uniform. Configurations where one component localizes into a central core while the other segregates to the periphery are labeled as core-rim (concentric-ring) states. Finally, when the density distribution breaks continuous rotational symmetry, forming multiple lobes or localized peaks along the rim, we classify the state as azimuthally modulated.

\subsection{Axial confinement vs. population imbalance}

\begin{figure}[t] 
	\begin{minipage}[b]{0.5\textwidth}
		\begin{overpic}[width=1\linewidth]{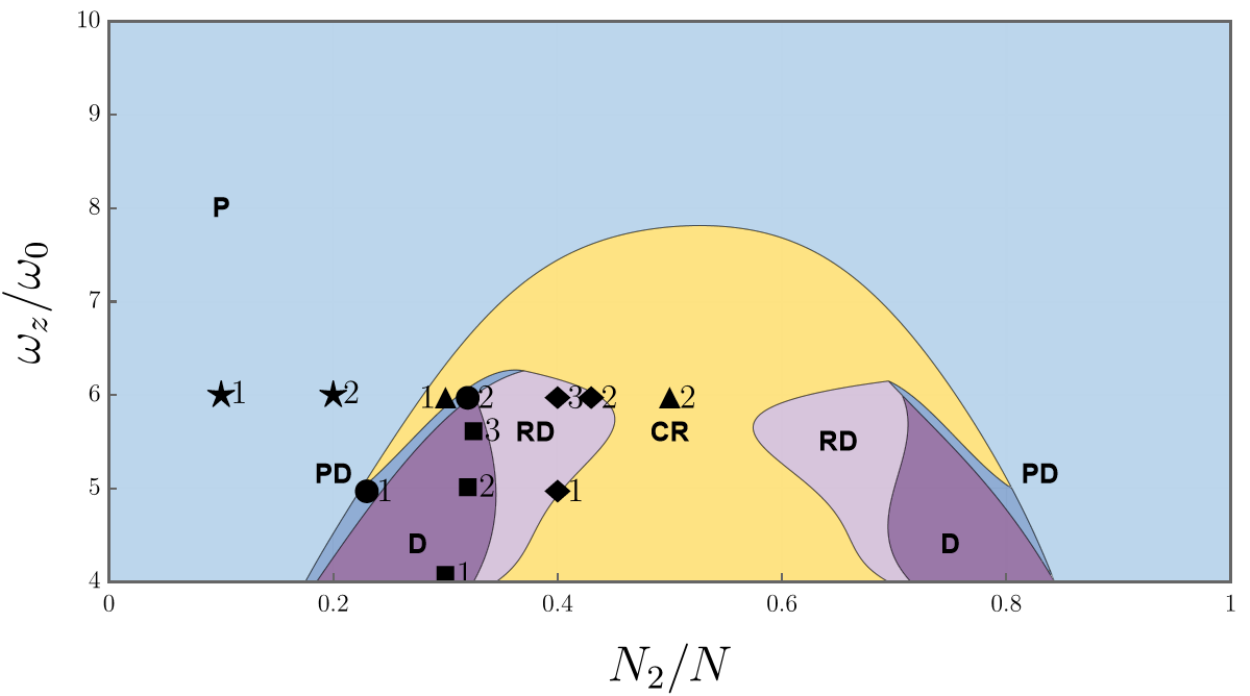} 
			\put(0,53){\small\bfseries (a)} 
		\end{overpic} 
	\end{minipage} 
	
	\vspace{6pt}
	
	\begin{minipage}[b]{0.5\textwidth}
		\begin{overpic}[width=1\linewidth]{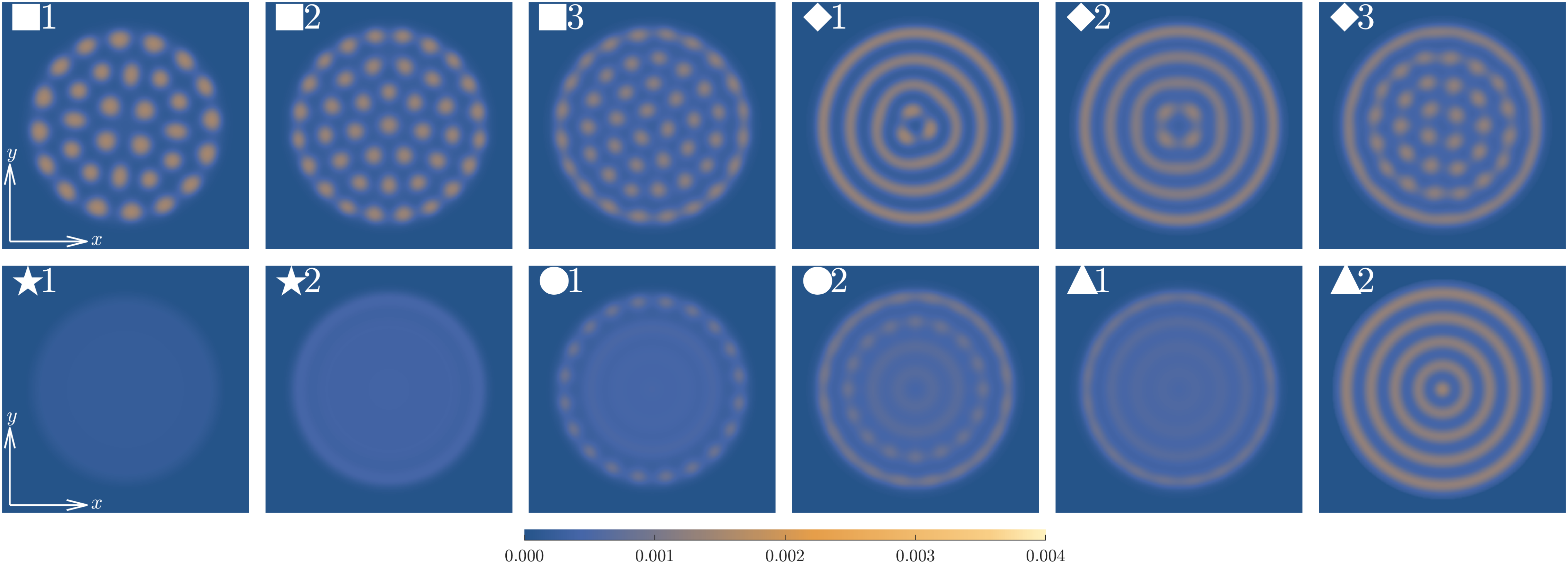}
			\put(0,37){\small\bfseries (b)}
		\end{overpic}
	\end{minipage}
	\caption{\label{fig:Phasediagram_omega}
		(a) Phase diagram in the $(\omega_z/\omega_0, N_2/N)$ plane for 
		$\gamma = 1$, $a_{11} = 100 a_0$, $a_{22} = 90 a_0$, $a_{12} = a_{21} = 95 a_0$, 
		$R_0 = 20$ $\mu$m, $V_0 = 100 \hbar \omega_0$, $\mu_{1}=6\mu_{B}$, $\mu_{2}=-6\mu_{B}$, and total atom number $N = 5 \times 10^{5}$. Colored regions indicate distinct density morphologies: pancake (P), pancake-droplet (PD), droplet (D), ring-droplet (RD), and concentric rings (CR).
		(b) Column-density profiles $n_{2}(\bm{\rho})$ at selected points in (a), labelled by markers. The field of view is $44.61 \times 44.61~\mu\mathrm{m}^2$. Density is plotted in units of $N_2 \ell^{-2}$.
	}
\end{figure}

We first examine how the interplay between the axial confinement strength and the population fraction $N_2/N$ (at fixed total atom number $N = N_1 + N_2$) governs the ground-state morphology. In this subsection, the contact interaction strengths are fixed at
\(
a_{11} = 100 a_0,\;
a_{22} = 90 a_0,\;
a_{12} = a_{21} = 95 a_0,
\)
ensuring only a slight asymmetry in the short-range interactions. 
The condition $a_{11}\neq a_{22}$ should be viewed as a tunable effective interaction asymmetry, rather than as a strict microscopic requirement. In multicomponent dipolar gases, the scattering lengths generally depend on the chosen internal states and can be adjusted using magnetic Feshbach resonances~\cite{Pfau2005PRLFeshbach,Ferlaino2020PRA}. By preparing the two components in different internal or dressed states and tuning the relevant scattering channels, one can therefore realize different effective intraspecies interactions in principle. The values used here, $a_{11}=100a_0$ and $a_{22}=90a_0$, differ by only 10\% and are intended to represent a weak state-dependent contact asymmetry. A direct implementation with antiparallel chromium Zeeman states may be limited by dipolar relaxation or spin-changing losses. Similar effective interaction settings, however, could be explored through dressed-state engineering, rotating-field control of the dipolar interaction~\cite{Pfau2002PRL,Benjamin2018PRL}, or heteronuclear dipolar mixtures~\cite{Ferlaino2018PRL}. Thus, the conclusions of this work rely mainly on the presence of a controllable interaction imbalance, rather than on the microscopic origin of the specific values of $a_{11}$ and $a_{22}$.
By scanning the axial frequency \(\omega_z\) and the population ratio \(N_2/N\) over experimentally relevant ranges, we determine the stationary states via energy minimization. The resulting phase diagram is presented in \reffig{fig:Phasediagram_omega}(a). 

A salient feature of \reffig{fig:Phasediagram_omega}(a) is the approximate mirror symmetry about the line \(N_2/N = 0.5\). This symmetry arises because both components are ${}^{52}\mathrm{Cr}$, sharing identical mass and absolute dipole moments; the slight deviation stems from the minor inequality in intra-component scattering lengths (\(a_{22} \neq a_{11}\)). This observation establishes population imbalance as the primary control parameter: for a fixed $\omega_z$, deviating from equal populations drives the system through an identical sequence of morphological transitions on either side of \(N_2/N = 0.5\), with only minor shifts in the phase boundaries. \reffigure{fig:Phasediagram_omega}(b) displays representative column-density distributions for the five identified morphologies: pancake (P), pancake-droplet coexistence (PD), droplets (D), ring-droplet coexistence (RD), and concentric rings (CR). Here the term ``droplet'' refers to a trapped density-peak domain within the finite box, rather than to a self-bound quantum droplet stabilized by beyond-mean-field effects in free space.

\paragraph*{Highly imbalanced populations and strong axial confinement.} 
In the regime of strong population imbalance (\(N_2/N \gtrsim 0.7\) or \(N_2/N \lesssim 0.3\)), the majority component dominates the mean-field energy, suppressing symmetry breaking. Consequently, the ground state remains a nearly uniform, disk-shaped cloud (P phase). A similar suppression of modulation occurs under strong axial confinement (\(\omega_z \gtrsim 6 \omega_0 \)). Here, the tight confinement enhances the effective two-dimensional interaction scale, rendering the uniform state energetically favorable. Near the P–PD boundary, a weak precursor ring may emerge at the periphery, but the bulk density remains largely unmodulated.

\paragraph*{Intermediate confinement and droplet formation.}
For moderate axial frequencies ($\omega_z \lesssim 6 \omega_0$), the axial width increases, reducing the effective two-dimensional contact energy scale and modifying the momentum dependence of the dipolar kernel. As a result, radial and azimuthal density modulations become energetically competitive. In this regime, we observe a cascade of droplet-bearing states (PD, D, and RD). The PD phase is characterized by a smooth bulk surrounded by a ``necklace" of regularly spaced droplets at the boundary, reflecting a balance between short-range repulsion, anisotropic DDI, and the boundary constraint imposed by the finite circular box.
Deeper into the droplet (D) region, the density fully fragments into localized peaks separated by a low-density background. Decreasing $\omega_z$ generally increases the droplet size and contrast while reducing their number. In dense droplet arrays (D and partial RD), the outermost droplets align conformally with the circular boundary, while the inner droplets adopt irregular, domain-like arrangements characteristic of dipolar media.

\paragraph*{Ring-droplet coexistence and concentric rings.} 
The RD phase represents a hybrid state where localized droplets near the center coexist with smooth rings at the periphery. This texture implies the simultaneous activation of multiple length scales in the dipolar interaction: the core favors localization to minimize inter-component overlap, while the boundary curvature stabilizes extended ring-like domains. Approaching equal populations (\(N_2/N \approx 0.5\)), the spatial overlaps and mean fields of the two components become comparable. The system minimizes energy by organizing into alternating concentric rings (CR) of nearly equal width. Unlike the transient rings near the P-PD boundary, these structures are robust, periodic, and preserve global rotational symmetry.

\subsection{Contact-interaction imbalance vs. population imbalance}

\begin{figure}[t]
	\begin{minipage}[b]{0.5\textwidth}
		\begin{overpic}[width=1\textwidth]{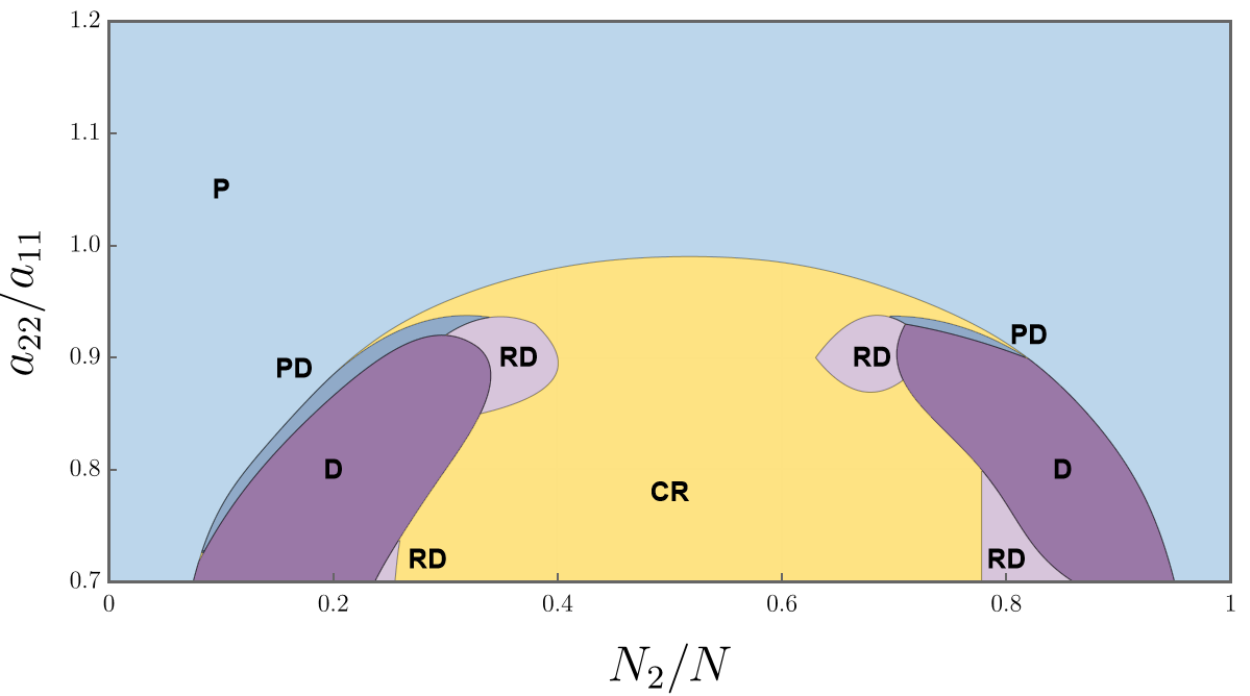}
		\end{overpic}
	\end{minipage}
	\caption{\label{fig:Phasediagram_a22}
		Phase diagram in the $(a_{22}/a_{11},\, N_2/N)$ plane for
		$\gamma = 1$, $a_{11} = 100 a_{0}$, $a_{12} = a_{21} = 95 a_{0}$,
		$R_0 = 20~\mu\mathrm{m}$, $V_0 = 100 \hbar \omega_0$,
		$\mu_{1} = 6\mu_{B}$, $\mu_{2} = -6\mu_{B}$,
		$\omega_z = 5 \omega_0$, and total atom number
		$N = 5 \times 10^{5}$. The color coding follows \reffig{fig:Phasediagram_omega}.
	}
\end{figure}

To test the robustness of the pattern sequence, we investigate the influence of contact-interaction imbalance by constructing a phase diagram in the $(a_{22}/a_{11},\, N_2/N)$ plane (see \reffig{fig:Phasediagram_a22}). Here, the ratio $a_{22}/a_{11}$ acts as a tuning knob for the relative "softness" of the components, while keeping $a_{12}$ and dipolar parameters fixed. The resulting diagram reproduces the same five morphologies observed in the $\omega_z$ scan in \reffig{fig:Phasediagram_omega}, confirming that the pattern selection is robust against the specific microscopic details of the interaction tuning.

Consistent with the $\omega_z$ scan, the P phase dominates at extreme population imbalances ($N_2/N \ll 0.5$ or $\gg 0.5$), where the minority component lacks sufficient density to drive symmetry breaking. Similarly, near the symmetric interaction point ($a_{22}/a_{11} \simeq 1$), the stability criterion $g_{11} g_{22} \gtrsim g_{12}^2$ is satisfied, suppressing demixing even in the presence of DDI.
For intermediate contact-interaction imbalance ($a_{22}/a_{11} \leq 0.95$), the PD, D, and RD phases emerge. Notably, these regions are asymmetric about $N_2/N = 0.5$. When $a_{22} < a_{11}$, component 2 is more compressible. Consequently, modulated phases are more prevalent when component 2 is the minority ($N_2/N < 0.5$), as it can be easily squeezed by the stiffer majority component and the box boundary into droplets or rings.

The apparent fragmentation of the RD sector near $a_{22}/a_{11}\simeq 0.8$ should not be interpreted as a separate phase. Rather, it reflects the projection of a multidimensional morphology boundary onto the two-parameter plane shown in \reffig{fig:Phasediagram_a22}, where small changes in the energy balance can favor either PD- or D-like local minima over the coexistence texture.
Collectively, \reffig{fig:Phasediagram_omega} and \reffig{fig:Phasediagram_a22} demonstrate that the population ratio $N_2/N$ is the dominant parameter determining the topology of the ground state, while other parameters ($\omega_z$, $a_{ij}$) primarily shift the boundaries for the onset of modulation.

\subsection{Structural correspondence to microphase-separated states}

It is instructive to compare the morphologies identified in \reffig{fig:Phasediagram_omega} and \reffig{fig:Phasediagram_a22} with the standard sequence of microphase-separated states known from diblock-copolymer systems. In the absence of dipolar interactions ($\gamma=0$), the binary condensate is expected to display either a miscible configuration or macroscopic phase separation, depending on the competition between intra- and intercomponent contact interactions. In contrast, the modulated states found here only emerge when the dipolar interaction is sufficiently strong. For example, for the representative point $(\omega_z/\omega_0, N_2/N)=(5,0.3)$ in \reffig{fig:Phasediagram_omega}(a), pronounced density modulation appears only when $\gamma$ is close to unity. This shows that the observed morphologies originate from the competition between local interactions, nonlocal dipolar coupling, and the finite confining geometry.

Within this perspective, the population ratio $N_2/N$ plays a role analogous to an effective volume fraction in classical microphase separation, in the sense that it primarily controls the topology of the stationary pattern. Near $N_2/N \approx 0.5$, where the two components have comparable populations, the system favors connected alternating domains and forms concentric rings. As the population ratio moves away from balance, maintaining a continuous ring topology becomes increasingly costly for the minority component, and the pattern evolves toward disconnected density peaks. This behavior is qualitatively consistent with the progression from lamellar-like to cylindrical-like morphologies in block-copolymer systems.

The correspondence, however, should be understood at the level of morphology and length-scale selection rather than as a literal identity of the underlying free-energy functionals. In particular, the circular hard-wall-like box imposes a strong geometric constraint absent in bulk copolymer systems. As a result, stripe-like modulations near balanced populations are replaced by concentric rings that conform to the boundary, while droplet arrays remain compatible with the finite domain over a broader range of imbalance. In this sense, the finite dipolar condensate provides a realization of microphase-pattern selection under nonlocal interactions and geometric frustration.
 
\subsection{Finite-size coexistence and geometric frustration}

The finite circular box introduces effects that are absent in bulk systems, most notably geometric frustration and the coexistence of distinct morphologies within a single confined domain. In particular, the PD and RD regimes can be viewed as finite-size coexistence states that interpolate between the limiting topologies. In an extended system, transitions between connected and disconnected patterns may occur more sharply, whereas in a mesoscopic box the boundary and the finite number of available modes allow different local structures to occupy different radial regions. This naturally gives rise to mixed textures, such as droplets in the interior coexisting with ring-like order near the boundary.

A second consequence of the finite geometry is the preference for concentric rings over straight stripe-like modulations near balanced populations. In a circular box, straight stripes would intersect the hard-wall boundary and generate a large mismatch between the preferred modulation pattern and the confining geometry. By contrast, concentric rings adapt naturally to the rotational symmetry of the trap and minimize boundary frustration, even though their curvature introduces an additional energetic cost. The resulting ring states should therefore be understood as geometry-selected analogs of lamellar ordering in a finite domain rather than as direct counterparts of bulk stripe phases.

These observations emphasize that the stationary patterns found here arise not only from the competition between local and nonlocal interactions, but also from the discrete mode structure imposed by the finite box. The morphology is thus selected by the interplay of intrinsic wavelength preference and geometric commensurability.

\section{Structure Factor And Scaling Analysis}
\label{SEC:structure factor and scaling analysis}

We now analyze the modulated states in momentum space in order to identify their characteristic length scale and to test whether this length scale is intrinsic to the interaction balance rather than imposed by the box size. The emergence of a nonzero peak in the structure factor provides a direct diagnostic of microphase formation. We then examine how the corresponding real-space spacing varies with the axial confinement length and show that it follows a robust confinement-controlled scaling, with additional finite-size steps arising from geometric commensurability.

\subsection{Structure factor and evidence for microphase separation}

\begin{figure}[t]
	\begin{minipage}[b]{\linewidth}
		\begin{overpic}[width=1\textwidth]{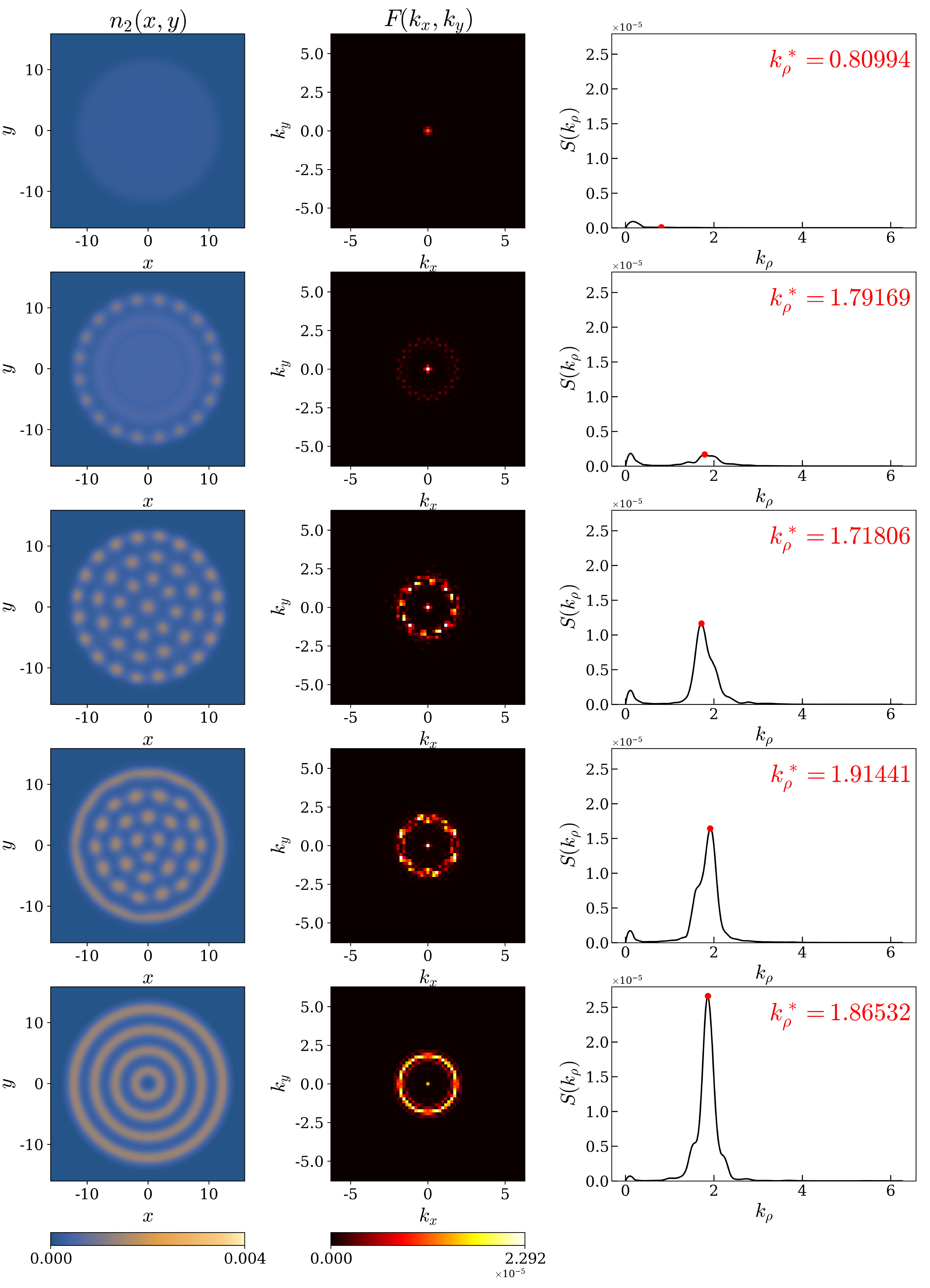}
		\end{overpic}
	\end{minipage}
	\caption{\label{fig:structure_factor_20}
		(Color online) Characterization of microphase separation. 
		First column: Real-space density distribution. 
		Second column: Two-dimensional Fourier power spectrum $F(k_x, k_y)$ in the range $k_{x,y} \in [-2\pi/\ell, 2\pi/\ell]$. The FFT satisfies Parseval's theorem: $\int F(\bm{k}_{\bm{\rho}}) d\bm{k}_{\bm{\rho}} = \int |\delta n_2(\bm{\rho})|^2 d\bm{\rho}$. 
		Third column: One-dimensional radial structure factor $S(k_\rho)$. 
		The five rows correspond to the representative points marked in \reffig{fig:Phasediagram_omega}(a), with parameters ($\omega_z / \omega_0$, $N_2/N$) = ($5$, $0.10$), ($5$, $0.24$), ($5$, 0.30), ($5$, $0.38$), ($5$, $0.50$) from top to bottom.
	}
\end{figure}

\begin{table*}[t]
	\begin{ruledtabular}
		\caption{Classification of phases based on structural signatures. \label{Table:structure factor}}
		\begin{tabular}{l c c c} 
			\textbf{Phase}      
			& \textbf{Structure factor $F(k_x, k_y)$}
			& \textbf{Radial profile $S(k_{\rho})$} 
			& \textbf{Physical Analogue} \\
			\midrule
			P (Pancake) & 
			Central peak at $\bm{k}=0$ & 
			Monotonic decay or low- $k$ peak & 
			Uniform fluid \\
			CR (Concentric Rings) & 
			Sharp, continuous ring at $|\bm{k}|=k^*$ & 
			Sharp single peak at $k_{\rho}=k_{\rho}^{*}$ & Smectic/Lamellar-like \\
			D (Droplet) & 
			Discrete hexagonal Bragg peaks & 
			Single peak at $k_{\rho}=k_{\rho}^{*}$ & 
			Triangular crystal/Cylindrical \\
			RD (Coexistence) & 	
			Diffuse ring + discrete peaks & 
			Broadened peak distribution & 
			Glassy/Multi-domain \\
		\end{tabular}
	\end{ruledtabular}
\end{table*}

To quantify the translational symmetry breaking, we analyze the static structure factor for five representative states identified in the phase diagram [see \reffig{fig:Phasediagram_omega}(a)]. We define the two-dimensional Fourier power spectrum of the density fluctuations as:
\begin{equation}
	F(k_x, k_y) = F(\bm{k}_{\bm{\rho}}) = \frac{1}{(2 \pi)^2} \left| \int d\bm{\rho} \delta n_2(\bm{\rho}) e^{-i \bm{k}_{\bm{\rho}} \cdot \bm{\rho}} \right|^{2},
\end{equation}
where $\bm{k}_{\bm{\rho}}=(k_x, k_y)$ is the in-plane wave vector. The fluctuation is given by $\delta n_2(\bm{\rho}) = n_2(\bm{\rho}) - \langle n_2 \rangle$, with $\langle n_2 \rangle = (\pi R_{0}^{2})^{-1} \int d\bm{\rho}' n_2(\bm{\rho}')$ denoting the spatial average.
To identify the dominant length scale, we compute the azimuthally averaged radial structure factor:
\begin{equation}
	S(k_{\rho}) = \frac{1}{2 \pi} \int_{0}^{2 \pi}  F(\bm{k}_{\bm{\rho}}) k_{\rho} dk_{\theta},
\end{equation}
where $(k{\rho}, k_{\theta})$ are the polar coordinates in momentum space.

The results are displayed in \reffig{fig:structure_factor_20}, and the spectral signatures of the different phases are summarized in \reftable{Table:structure factor}. A crucial observation is that for all modulated phases (CR, D, RD), the power spectrum is concentrated around a nonzero momentum shell $k_{\rho}^{*} > 0$. This defines a characteristic wavelength $\lambda^* = 2\pi / k_{\rho}^{*}$ in real space, corresponding to the periodicity of the concentric rings or the lattice constant of the droplet array.
Notably, the peak intensity of $S(k_{\rho})$ increases as $N_2/N$ approaches $0.5$, reflecting the deepening of the modulation amplitude near the symmetric filling, reminiscent of stronger segregation in block-copolymer morphologies.

\begin{figure}[t]
	\begin{minipage}[l]{0.49\textwidth}
		\begin{overpic}[width=1\textwidth]{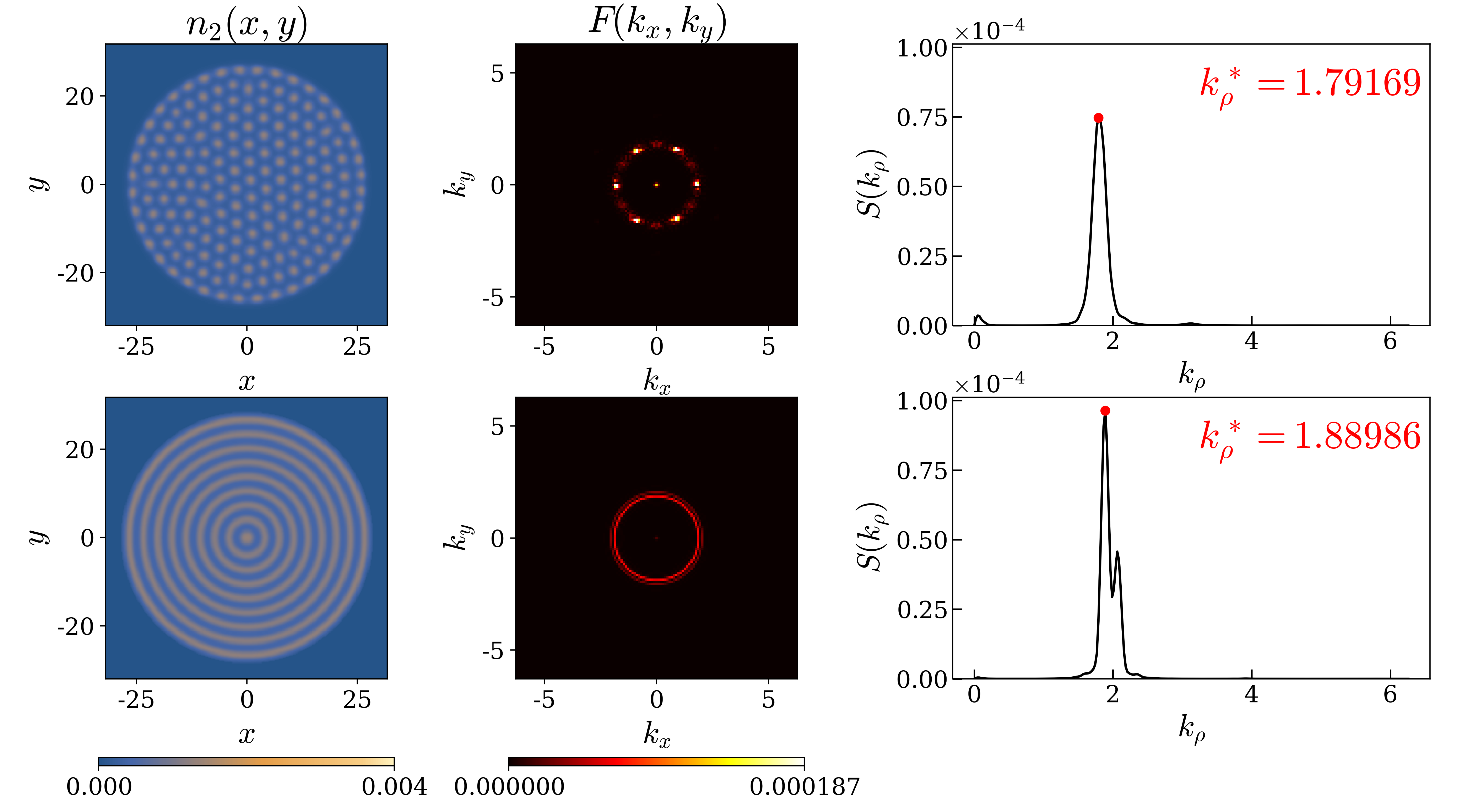}
		\end{overpic}
	\end{minipage}
	\caption{\label{fig:structure_factor_40}
		Robustness of the characteristic scale against system size. Comparison of real-space density, Fourier spectrum, and radial structure factor for a larger box radius $R_0 = 40$ $\mu$m. Parameters: $\omega_z = 5 \omega_0$, $N = 2 \times 10^{6}$, with $N_2/N = 0.32$ (Droplets) and $0.5$ (Rings). The selected $k^*$ remains consistent with the $R_0=20$ $\mu$m case in \reffig{fig:structure_factor_20}.
	}
\end{figure}

To confirm that this length scale is intrinsic and not an artifact of the boundary geometry, we compare the results for box radii $R_0 = 20~\mu\mathrm{m}$ (\reffig{fig:structure_factor_20}) and $R_0 = 40~\mu\mathrm{m}$ (\reffig{fig:structure_factor_40}). When increasing $R_0$ from $20\,\mu{\rm m}$ to $40\,\mu{\rm m}$, we scale the atom number from $5\times10^5$ to $2\times10^6$ so that the mean areal density is kept approximately fixed. We find that the characteristic wave vector $k_{\rho}^{*}$ is invariant with respect to the box size $R_0$. This independence indicates that the mode selection arises from the competition between the internal kinetic and interaction energies—specifically the momentum dependence of the dipolar kernel—rather than from geometric excitation modes of the box.

\subsection{Characteristic length scale and confinement-controlled scaling}

Having established that the modulated phases are associated with an intrinsic nonzero wave vector, we now examine how the corresponding real-space spacing depends on the axial confinement. In the quasi-2D regime, 
\begin{equation}
	\Psi_i(\mathbf{r})=\psi_i(\boldsymbol{\rho})\frac{1}{(\pi l_z^2)^{1/4}}
	e^{-z^2/(2l_z^2)},
\end{equation}
where $l_z=\sqrt{\hbar/(m\omega_z)}$ is the axial confinement length. After integrating over the confined direction, both the contact and dipolar interaction terms acquire an explicit dependence on $l_z$, so that the effective in-plane interactions are controlled by the transverse thickness of the condensate.

\begin{figure}[t]
	\begin{minipage}[l]{0.49\textwidth}
		\begin{overpic}[width=1\textwidth]{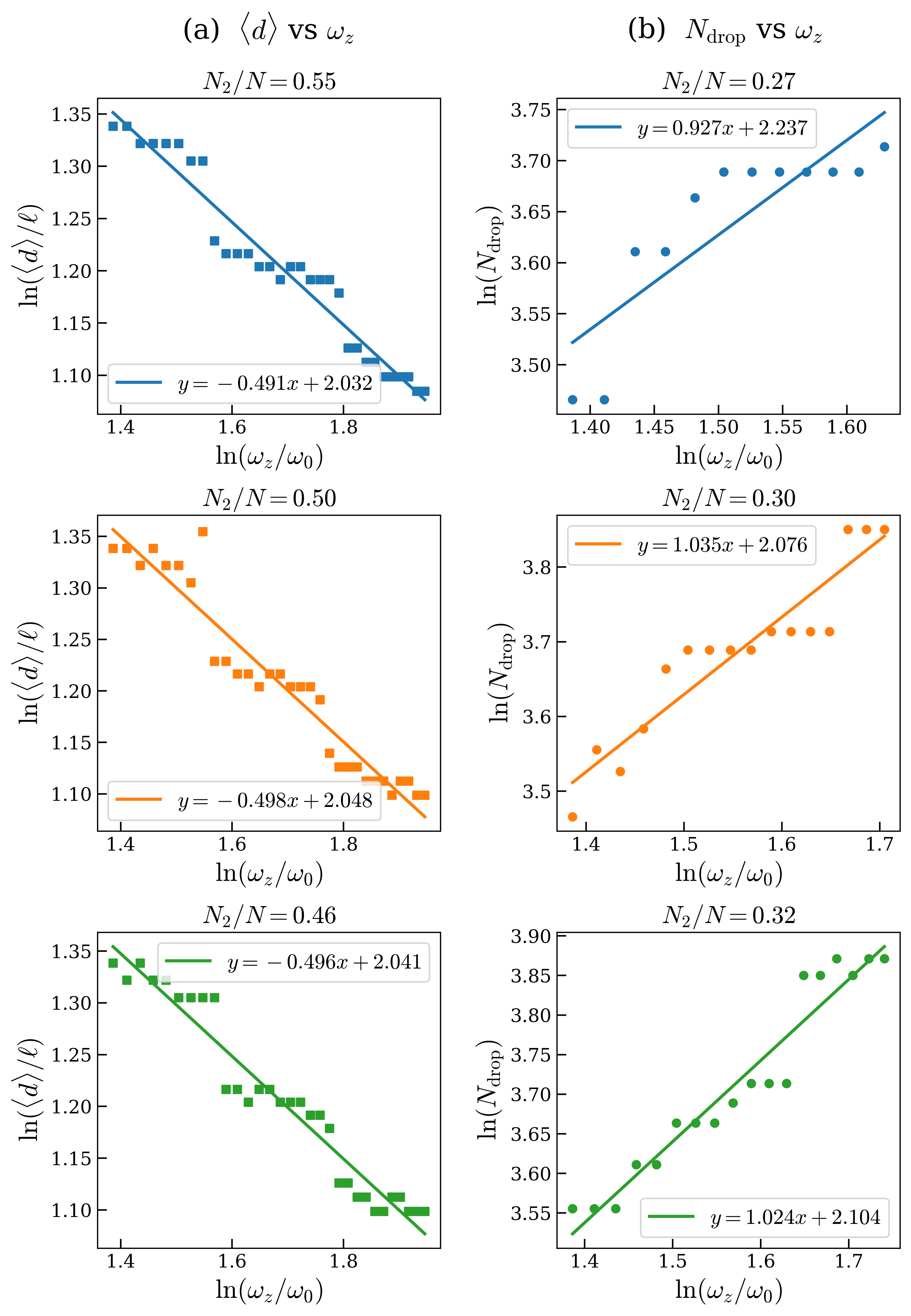}
		\end{overpic}
	\end{minipage}
	\caption{\label{fig:scaling law}
		Confinement-controlled scaling and geometric frustration. (a) Log-log plot of the mean inter-ring spacing $\langle d \rangle$ (in units of $\ell$) versus axial frequency $\omega_z$ (in units of $\omega_0$). The solid line indicates the power-law fit $\langle d \rangle \propto \omega_z^{-0.49}$. (b) Log-log plot of the droplet number $N_{\text{drop}}$ versus $\omega_z$. The step-like features indicate 
		locking of the droplet number due to the finite box geometry.
	}
\end{figure}

To quantify the length scale of the resulting pattern, we define a characteristic spacing $d$ from the modulated density profiles. For concentric-ring states, $d$ is taken as the mean radial distance between neighboring rings. For droplet states, it is estimated from the average droplet spacing derived from the droplet number within the box. \reffigure{fig:scaling law} shows that, over the parameter range explored here, the characteristic spacing follows a robust power-law dependence on the axial frequency,
\begin{equation}
	d \propto \omega_z^{-1/2},
\end{equation}
or equivalently
\begin{equation}
	d \propto l_z.
\end{equation}
This scaling indicates that the in-plane modulation length is directly controlled by the confinement scale in the  $z$-direction, reflecting a confinement-induced dimensional crossover in the effective interactions.

An additional feature of \reffig{fig:scaling law} is the appearance of step-like plateaus around the mean scaling trend. These steps arise from the finite circular geometry. Because the number of rings or droplets in the box must remain an integer, the system cannot continuously adjust its topology as the preferred spacing changes with $\omega_z$. Instead, it remains locked within a given distinct patterns over a finite parameter interval and then undergoes a discrete rearrangement when the mismatch becomes sufficiently large. The observed staircase structure therefore reflects geometric frustration and commensurability effects in a finite domain superimposed on the underlying confinement-controlled scaling. Equivalently, the dimensionless product $k^*l_z$ remains approximately constant over the scanned range, apart from finite-size jumps associated with changes in the number of rings or droplets.

\section{Characterization of the modulation amplitude} \label{SEC:the modulated-unmodulated transition}

To quantify the strength of the density modulation, we introduce the global density contrast of the minority component,
\begin{equation}
	C_2=
	\frac{\max[n_2(\boldsymbol{\rho})]-\min[n_2(\boldsymbol{\rho})]}
	{\max[n_2(\boldsymbol{\rho})]+\min[n_2(\boldsymbol{\rho})]},
\end{equation}
where $n_2(\boldsymbol{\rho})$ is the corresponding column density within the box. By construction, $C_2$ provides a scalar measure of the modulation amplitude: $C_2\to 0$ corresponds to a nearly uniform state, whereas larger values of $C_2$ indicate increasingly pronounced density modulation.

\begin{figure}[t]
	\begin{minipage}[b]{0.49\textwidth}
	\begin{overpic}[width=\linewidth]{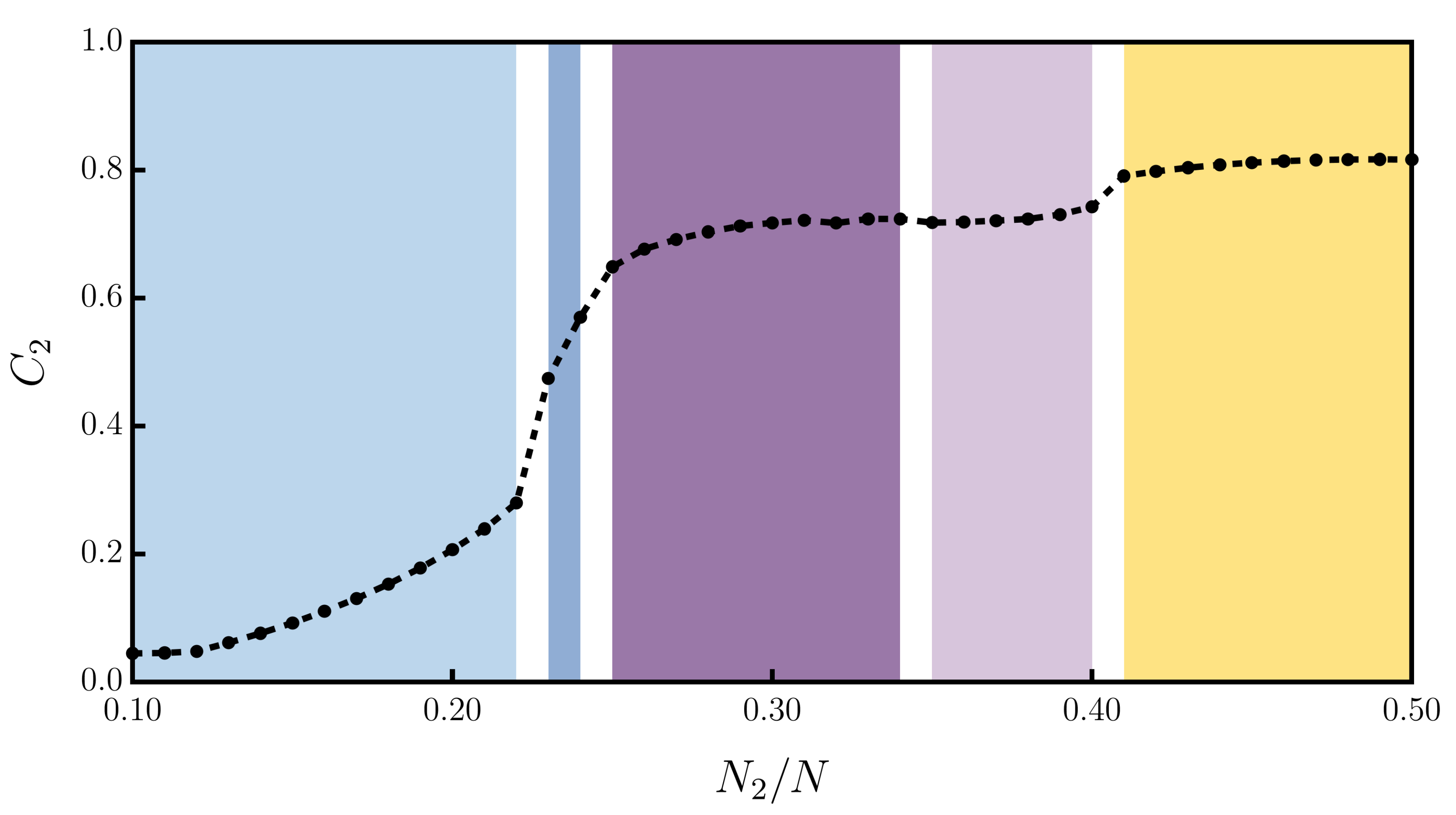}
		\put(-220,135){\small\bfseries (a)}
	\end{overpic}
	\end{minipage}
	\hfill
	\begin{minipage}[b]{0.49\textwidth}
	\begin{overpic}[width=\linewidth]{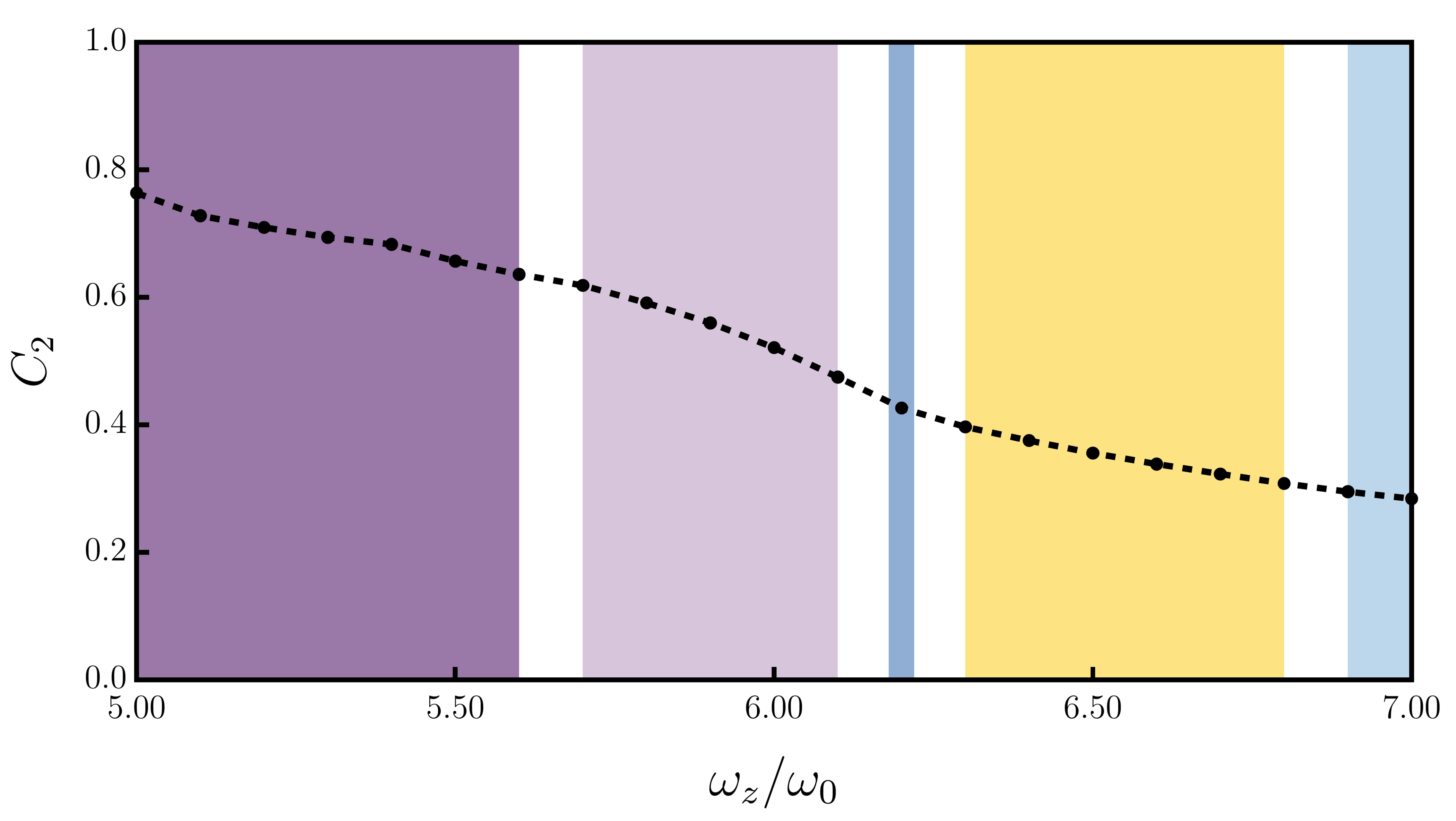}
		\put(-220,135){\small\bfseries (b)}
	\end{overpic}
\end{minipage}
	\caption{(Color online) Characterization of the modulation amplitude.
		(a) Density contrast $C_{2}$ versus population ratio $N_2/N$ at fixed $\omega_z = 5 \omega_0$. The kinks in the slope indicate the onset of new dominant Fourier modes corresponding to the phase boundaries in \reffig{fig:Phasediagram_omega}.
		(b) Density contrast $C_{2}$ versus axial confinement $\omega_z$ at fixed $N_2/N = 0.34$. The smooth decay illustrates the suppression of the density modulation by tight confinement.}
	\label{fig:C2_phasetransition}
\end{figure}

\reffigure{fig:C2_phasetransition}(a) shows $C_2$ as a function of the population ratio $N_2/N$ at fixed axial confinement. The contrast evolves smoothly across the phase diagram, while clear changes in slope correlate with the morphology boundaries identified in \reffig{fig:Phasediagram_omega}. In a finite mesoscopic box, such behavior is naturally interpreted as a finite-size--rounded crossover in a global scalar observable rather than as a sharp thermodynamic singularity. The continuous variation of $C_2$ therefore reflects the gradual growth of the modulation amplitude, even though the underlying pattern topology may change between ringlike, droplet-like, and coexistence textures.

\reffigure{fig:C2_phasetransition}(b) displays $C_2$ as a function of $\omega_z$ at fixed $N_2/N$. As the confinement becomes tighter, the modulation amplitude decreases monotonically, indicating that strong axial confinement suppresses the in-plane density modulation. Comparing \reffig{fig:C2_phasetransition}(a) and \reffig{fig:C2_phasetransition}(b), we find that $C_2$ is more sensitive to the population imbalance than to the confinement strength. This further supports the conclusion that the population ratio acts as the primary control parameter for topology selection, while the confinement mainly tunes the characteristic length scale and the modulation depth.

We emphasize that $C_2$ alone does not distinguish between different symmetry classes of modulated states. For example, both concentric-ring and droplet configurations may display comparable values of $C_2$. The identification of the underlying morphology therefore requires the complementary spectral analysis presented in Sec.~\ref{SEC:structure factor and scaling analysis}.

\section{Conclusion}
\label{SEC:Conclusion}

In this work, we have studied the stationary density patterns of a population-imbalanced binary dipolar Bose-Einstein condensate confined in a circular two-dimensional box. Within a mean-field framework, we systematically explored how the ground-state morphologies depend on the axial confinement $\omega_z$, the contact-interaction asymmetry $a_{22}/a_{11}$, and the population ratio. The resulting phase diagrams reveal a robust set of patterns: a nearly uniform \emph{pancake} state (P), \emph{pancake--droplet} coexistence (PD), triangular \emph{droplet} arrays (D), \emph{ring--droplet} coexistence (RD), and alternating \emph{concentric-ring} states (CR). The fact that similar morphology sequences appear under different tuning protocols shows that the pattern topology is controlled primarily by the population imbalance, while the confinement and interaction asymmetry mainly shift the modulation thresholds.

A key outcome of this work is the structural connection between the density patterns of the confined dipolar mixture and microphase-separated morphologies known from soft-matter systems. In this analogy, the atom-number ratio $N_2/N$ plays a role similar to an effective volume fraction: it selects whether the system favors connected ringlike domains or disconnected density peaks. The circular boundary further reshapes the bulk-like lamellar and cylindrical patterns into concentric rings and droplet arrays. This interpretation is supported by the structure-factor analysis, which shows that the modulated phases are characterized by an intrinsic nonzero wave vector.

We also found that the characteristic pattern spacing scales linearly with the axial confinement length,
\[
d \propto l_z \sim \omega_z^{-1/2}.
\]
This relation indicates that the transverse thickness of the condensate sets the effective in-plane length scale through the quasi-two-dimensional interactions. In a finite circular box, the smooth scaling trend is interrupted by discrete steps, reflecting geometric frustration and the integer locking of the number of rings or droplets.

The global contrast $C_2$ provides a convenient measure of the modulation amplitude across the phase diagram. Its smooth variation is consistent with finite-size rounding of the modulated--unmodulated crossover in a mesoscopic box. However, because $C_2$ does not distinguish between different pattern symmetries, the classification of ringlike and droplet-like states requires the complementary structure-factor diagnostics discussed above.

Several directions remain open for future work. Including Lee-Huang-Yang corrections would help clarify how beyond-mean-field stabilization competes with boundary-induced pattern selection in more strongly dipolar mixtures. Real-time simulations of parameter ramps could reveal the nucleation pathways, hysteresis, and defect formation associated with the morphology changes. From an experimental perspective, the antiparallel chromium configuration considered here should be regarded as an idealized minimal model, since dipolar relaxation may limit the lifetime in a direct implementation. Similar effective interactions may instead be engineered using rotating-field or microwave-dressing schemes, or realized in heteronuclear dipolar mixtures. Overall, our results show that box-trapped dipolar mixtures provide a useful platform for exploring finite-size pattern selection in nonlocal quantum fluids.

\begin{acknowledgments}
Z.W. was supported by NUAA (Grant No. 20241028700078W and No. 20251028700788X). W.B. was supported by NUAA (Grant No. 202510287091Z). J.-R.L. was supported by NUAA (Grant No. 20251028700803X). G.W. was supported by the National Natural Science Foundation of China (Grant No. 12375039). K.-T.X. was supported by the MOST of China (Grant No. G2022181023L) and NUAA (Grant No. YAT22005, No. 2023YJXGG-C32 and No. XCA2405004).
\end{acknowledgments}

\bibliography{references}

\end{document}